# Programmable Compute-in-Transit using Integrated Photonics

Imon Kundu, Livi Hammond, Jamie Todd, Kriti Goel, Peter Simpson, Jaganath Rajendra, Florent Michel, Jack Crawford, Flavio Bergamaschi, Robert Todd, Nick New
Optalysys
12B Platform, Leeds LS1 4JB, UK
imon.kundu@optalysys.com

***Abstract*— Modern hardware designs for AI and cryptography treat data transit and processing separately. At Optalysys we have demonstrated programmable Photonic hardware that computes mathematical functions on data that is in transit, enabling tens of GFLOPs of operations on data links.**

## I. Introduction

Data movement is increasingly the dominant bottleneck in modern computing systems, particularly for demanding workloads such as Artificial Intelligence (AI) [1], Post-Quantum Cryptography (PQC) [2], and Fully Homomorphic Encryption (FHE) [3]. Conventional architectures treat computation and communication as separate processes, with data transported to discrete processing units before operations can be applied. As system scale increases, this separation leads to inefficiencies in latency, bandwidth, and energy. Compute-in-Transit offers an alternative model, where computation is performed during data movement by embedding operations directly along the data path, shifting the focus from moving data to compute towards applying computation as data propagates through the system [4].

These workloads are characterised by extremely high computational demands, often reaching tera-scale to exa-scale operations per second. In AI systems, multiply-accumulate (MAC) operations on floating-point data dominate accelerator design, with modern hardware adopting reduced-precision formats to improve efficiency and reduce memory footprint. In contrast, cryptographic workloads such as FHE rely on structured polynomial transformations, including Fast Fourier Transforms (FFT) and Number Theoretic Transforms (NTTs), which form the computational backbone of secure processing. Although these domains differ in structure, both involve repeated application of regular operations over large data sets, making performance increasingly sensitive to data movement and memory access patterns. This is further amplified by the creation and propagation of intermediate data during computation, which must be repeatedly stored, transferred, and transformed across processing stages.

Digital Signal Processing (DSP) cores and Application-Specific Integrated Circuits (ASICs) are widely used to accelerate the dominant operations in these workloads, including MAC operations in AI and large-scale FFT and NTT computations in FHE [5], [6], [7]. However, scaling to very large transform sizes—such as 65,536-point NTTs required in certain FHE workloads or even larger ones in some Zero-Knowledge Proof schemes [8]—remains constrained by memory access patterns, interconnect complexity, and the cost of moving data between processing elements. In practice, these constraints limit overall system performance and highlight the growing imbalance between compute capability and data movement efficiency.

Conventional architectures treat data movement and computation as distinct stages, requiring data to be transferred across memory hierarchies and interconnects before processing. As workloads scale, this repeated movement of data, including intermediate results, introduces increasing overheads and limits achievable performance. Recent work has started to explore tighter integration between computation and data movement, including hardware–software co-design strategies and photonic implementations for structured workloads [9], [10], providing a foundation for architectures that more closely align computation with dataflow.

In this work, we present a prototype demonstration of the Compute-in-Transit architecture using a Photonic Integrated Circuit (PIC), enabling computation to be applied directly along the data path during transmission. The proposed implementation supports a range of mathematical operations through a programmable serial compute pipeline operating across high-speed transceiver lanes in a standard QSFP56 port. We demonstrate a prototype hardware platform with a custom driver IP that enables synchronous operation over asynchronous links. The prototype achieves sustained operation through integrated self-link fault detection and fully automated photonic calibration mechanisms.

## II. Pluggable Module for Compute-in-Transit

Our Pluggable Module is designed for computing data serially using Photonics. The PIC integrates Mach-Zehnder Modulators (MZMs) for data modulation; a proprietary passive diffractive device that computes Optical Fourier Transform (OFT) [9], [10], [11]; Photodiodes (PDs) in a balanced configuration for coherent detection; and Thermo-Optic Phase Shifters (PS) for phase control. Light from an external laser source is split into seven channels. Four of the seven channels are connected to MZMs. The output of each MZM is connected to a PS and split into two further channels – one of these interfaces with the OFT device and the other is used for coherent homodyne monitoring. Two of the seven channels are connected to PS directly. The output from each phase shifter is split into two, to form a total of four monitor channels. Each one of the monitor channels mix with the monitor outputs from the MZMs. The remaining seventh channel is forwarded to form a reference local oscillator (LO) channel to the receiver for coherent homodyne detection of the outputs from the OFT device. This reference channel is connected to a slow-speed Mach-Zehnder Interferometer (MZI) to control both phase and intensity. The intensity and phase-controlled output is split into four channels – each of these are connected to PS. The output from each PS is mixed with an output channel from the OFT device and is detected using coherent homodyne detection techniques. All coherent homodyne detection techniques used in the PIC use balanced PDs.

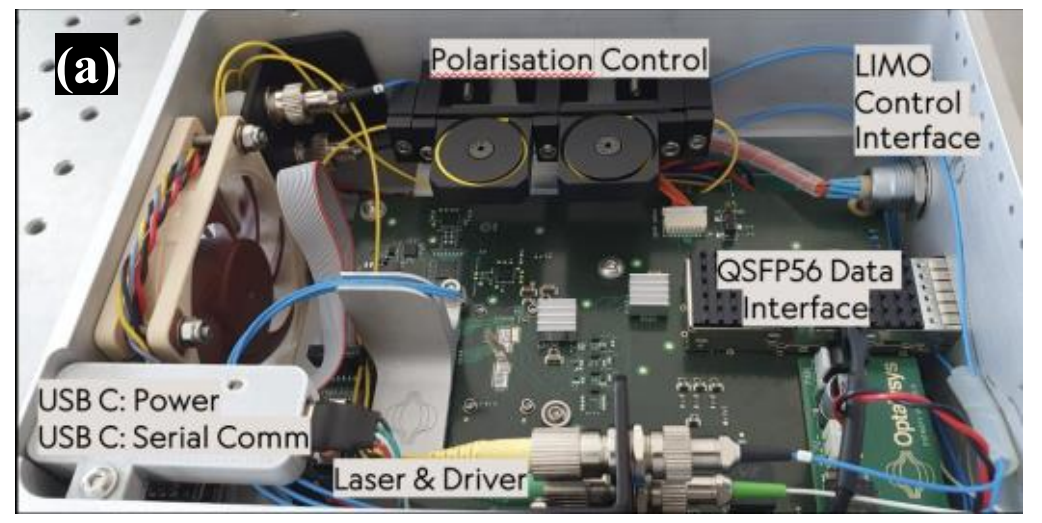

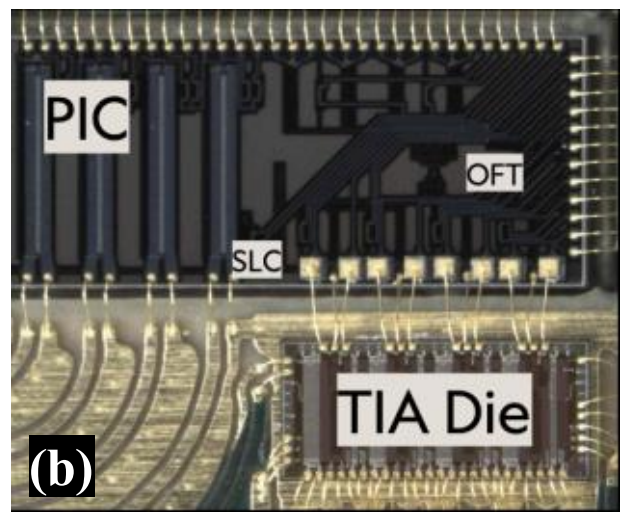

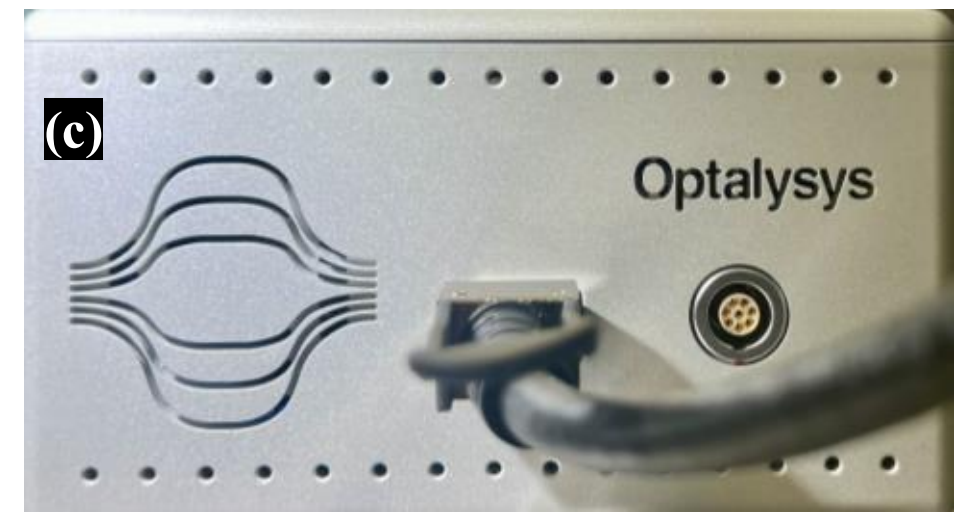

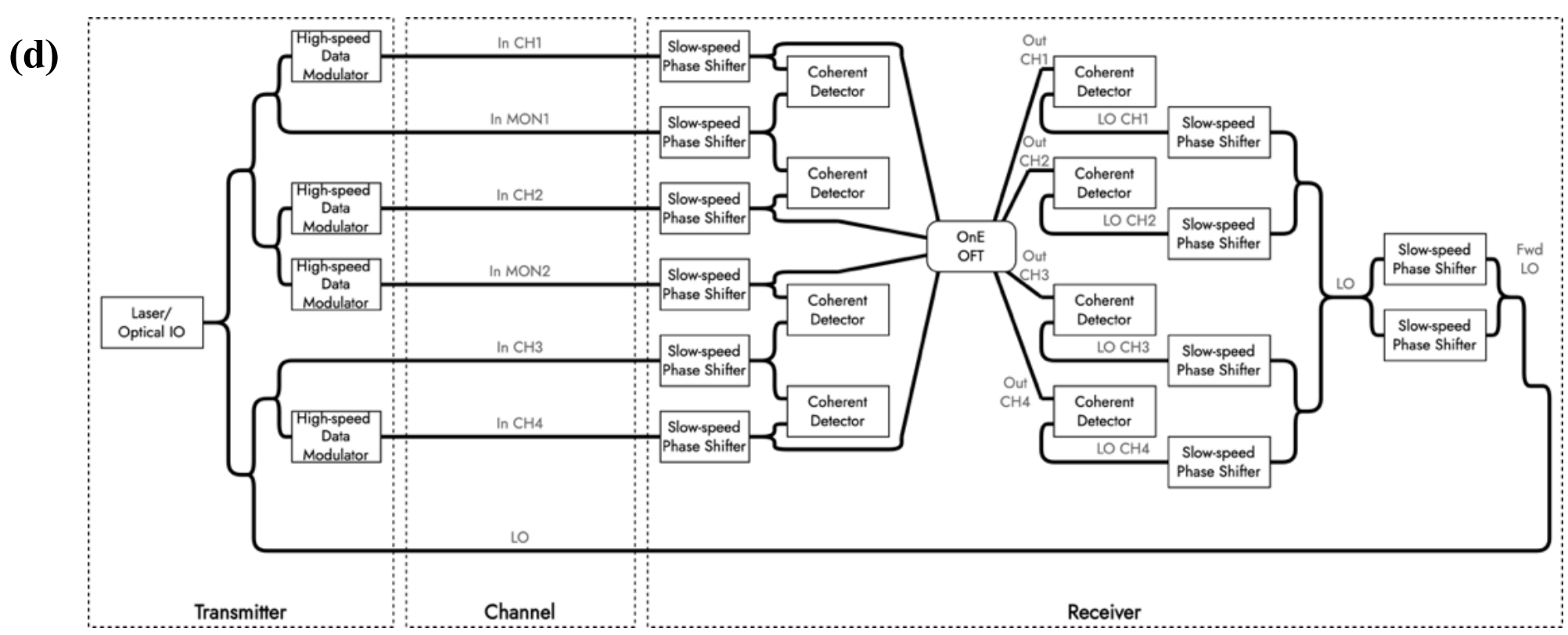

Fig. 1. (a) Internals of the Optalysys Pluggable Module. (b) Details of the PIC with the OFT device. The PDs in the PIC are packaged with Single Layer Capacitors (SLCs) and bonded with a TIA die. (c) Front view of the module. (d) Schematic of a phase balanced PIC with a passive diffractive device to perform optical Fourier transform. Coherent balanced photodiodes are used to detect partial results.

The PIC architecture [12] along with phase monitoring and compensation methods[13], data serialisation and compression methods [14], [15] is designed for serialised computation of OFT based computation on communication links [16]. The MZMs, PDs and PS used in the PICs are derived from foundry provided PDK cells, making our technology portable and scalable across different foundry platforms. The Photonic PDK devices used in the current system can operate at 14 GBaud/s (28 Gbps PAM4). The interface to the MZMs and PDs are Direct-Drive MZM Drivers and Trans-Impedance Amplifiers with a max baud rate of 56 GBaud/s (112 Gbps PAM4).

A QSFP56 port is used for data exchange between the module and a host FPGA (an AMD Alveo V80 Accelerator Card). The high-speed serial data from the QSFP56 TX lanes in the FPGA is used to drive the MZMs, and the multi-bit data from the PDs are sent via the QSFP56 RX lanes in the FPGA.

The module also integrates a microcontroller; laser and driver; polarisation control paddles; and power management system. An algorithm for automation of PIC signal integrity and initialisation logic is embedded into the microcontroller. The algorithm reacts to slow-speed time averaged sensor data from PDs using ADCs built in the microcontroller and sets PS using DACs in the microcontroller. The microcontroller also controls the configuration settings to the on-board Direct Drive devices. A USB-C serial port links the microcontroller to a host. Another USB-C port provides power to the module. The power consumption of the module is 9.7 W

## III. Computing-in-Transit Prototype

A Driver IP was developed and synthesised on the FPGA programmable logic to synchronise transmission, manage data exchange, compute results from the module, and manage backpressure between the module and application IPs. It also integrates a proprietary Error Correction IP. The Driver IP reads data from dedicated memory addresses, encodes, and sends/receives data to/from the Pluggable Module. The transmitted data to the module is restricted to binary (NRZ) levels. This is to compress the photonic computed result to PAM4 [9], [10]. The memory addresses to the Driver IP are also connected to a Network-on-Chip and can access High-Bandwidth Memory (HBM) in the FPGA. An application can access the memory addresses via PCIe using AMD AVED Management Interface. The Driver IP also integrates prototype modular arithmetic logic. These logic circuits implement novel, proprietary algorithms to compute modular arithmetic from OFT-based serial data.

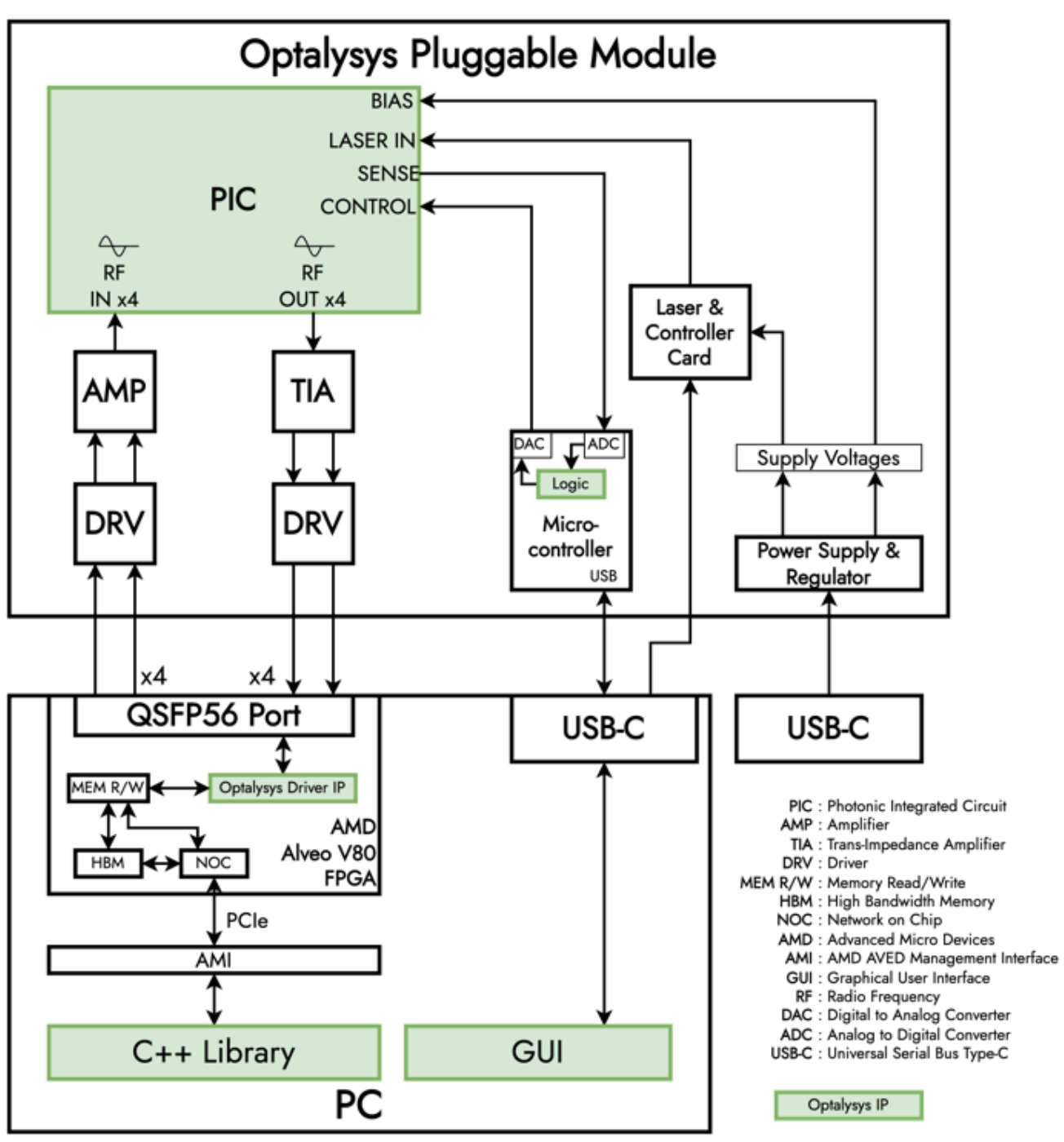

Fig. 2. Block diagram of the Compute-in-Transit prototype.

We have also developed a distributable C++ library that packages all system control functions for ease of developer integration. A Python based Graphical User Interface (GUI) was also developed to communicate with the microcontroller in the Pluggable Module to provide system control. The combination of the IPs, interfaces and Pluggable Module form a prototype Compute-in-Transit implementation. A block diagram of this system is presented in Fig 2.

The following operations are supported by this Compute-in-Transit system: 1xN FFT with Radix-4 decimation, 1 x N NTT with Radix-4 decimation, Addition & Subtraction, Multiplication, MAC, Modular Addition & Subtraction, Modular Multiplication, Modular MAC, Modular Multiplication & Subtraction (MSUB), Floating Point MAC. An example of programmable operation is shown in Fig 3.

Fig. 3. Execution log showing computation of a radix-4 NTT, modular addition and subtraction using the Compute-in-Transit prototype. The expected values indicate digitally equivalent computed using FPGA and difference indicate the deviation of the results obtained using Compute-in-Transit prototype from the FPGA computed results.

## IV. USE CASES: FULLY HOMOMORPHIC ENCRYPTION & POST QUANTUM CRYPTOGRAPHY

To demonstrate practical application use cases, we integrated the Compute-in-Transit prototype and its associated C++ library with Intel HEXL [17] for polynomial arithmetic acceleration and OpenFHE [18] for cryptographic primitives. This allowed us to evaluate the photonics compute engine within real lattice-based cryptographic workflows and verify correctness against established software implementations.

The first workload class we examined is CKKS-based fully homomorphic encryption [19]. CKKS supports approximate arithmetic over encrypted real-valued data and is commonly used in privacy-preserving computation and AI workloads. In practical deployments, ciphertexts accumulate noise as computations progress, eventually requiring a refresh operation, known as bootstrapping, to restore computational depth. Bootstrapping is dominated by repeated polynomial arithmetic, basis conversions, and large sequences of NTT/INTT operations, making it one of the primary computational and memory bottlenecks in FHE systems [20].

To evaluate the suitability of the Optalysys Compute-in-Transit Photonics engine for CKKS bootstrapping, we evaluated the OpenFHE iterative bootstrapping implementation using composite scaling, ring dimension 65536, and word size 32 bits. All NTT/INTT operations were executed on the photonic engine. The 65536-point NTTs were first decomposed into 256-point transforms and then further distributed across the four optical channels using a radix-4 decomposition. The complete bootstrapping workflow involved approximately 420 GBytes of streamed data passing through the photonic engine. Comparison against the native OpenFHE CPU implementation confirmed functional correctness of the optical computation.

The second application area we are investigating is ML-KEM [21], the lattice-based Key Encapsulation Mechanism standardized by NIST as FIPS 203. ML-KEM is one of the core building blocks of emerging post-quantum cryptographic protocols designed to resist attacks from future quantum computers. Similar to CKKS, ML-KEM relies heavily on polynomial arithmetic and repeated NTT/INTT operations, although with much smaller polynomial dimensions and moduli. The streamed architecture of the Optalysys Compute-in-Transit Photonics makes it well suited to these workloads despite the large difference in parameter sizes between CKKS and ML-KEM.

As a proof of concept, we are porting and integrating the mlkem-native implementation [22], developed within the Post-Quantum Cryptography Alliance (PQCA) ecosystem [23]. This integration will allow us to evaluate the suitability of the Compute-in-Transit Photonics architecture for lattice-based post-quantum cryptographic workloads and help characterize the performance of future photonic/optical acceleration systems for ML-KEM-style polynomial arithmetic.

## V. FUTURE WORK

We are integrating the compute-in-transit IPs in a monolithic silicon photonic chip to compute operations such as MAC, FFT and NTT during data transit. The technology integrates a 112 GBaud/s Photonic Integrated circuit (PIC) with Analog Mixed Signal (AMS), Analog Front End (AFE) electronics and custom high-speed logic operating at 28 GHz. This technology opens possibilities of enabling compute operations on standard transceiver links. The serial nature of the computation allows flexibility in bus width (4-bit to 64-bit). For example, 45 GFLOPS of FP4 (E2M1) MAC is enabled on a 112 GBaud/s photonic transceiver link for an additional 1 pJ/bit. Since the core compute technology is passive, the architecture also scales with the Silicon Photonics roadmap for higher data rates per lane.